 \documentclass[final,3p,times]{elsarticle}

\usepackage{graphicx}
\usepackage{amssymb}
\usepackage{subfig}
\usepackage{comment}
\usepackage{enumitem}
\usepackage{amsmath}
\usepackage{booktabs} 

\usepackage{lineno}
\usepackage{url}
\usepackage{xcolor}
\usepackage{soul}

\biboptions{sort&compress}

\begin{document}

\begin{frontmatter}

\title{
Twitter and Census Data Analytics to Explore Socioeconomic Factors for Post-COVID-19 Reopening Sentiment}





\author[label1,label4]{Md. Mokhlesur Rahman}
 \ead{mrahma12@uncc.edu}
\author[label2]{G. G. Md. Nawaz Ali\corref{cor1}}
 \ead{ggmdnawazali@ucwv.edu}
\author[label3]{Xue Jun Li}
 \ead{xuejun.li@aut.ac.nz}
\author[label1]{Kamal Chandra Paul}
\ead{kpaul9@uncc.edu}
\author[label3]{Peter H.J. Chong}
\ead{peter.chong@aut.ac.nz}
 \address[label1]{University of North Carolina at Charlotte, NC 28223, USA}
 \address[label2]{University of Charleston, WV 25304, USA}
 \address[label3]{Auckland University of Technology, Auckland 1010, NZ}
\address[label4]{Khulna University of Engineering \& Technology (KUET), Khulna 9203, Bangladesh}

 \cortext[cor1]{Corresponding author}

\begin{abstract}
Investigating and classifying sentiments of social media users (e.g., positive, negative) towards an item, situation, and system are very popular among the researchers. However, they rarely discuss the underlying socioeconomic factor associations for such sentiments. 
This study attempts to explore the factors associated with positive and negative sentiments of the people about reopening the economy, in the United States (US) amidst the COVID-19 global crisis. It takes into consideration the situational uncertainties (i.e., changes in work and travel pattern due to lockdown policies), economic downturn and associated trauma, and emotional factors such as depression. 
To understand the sentiment of the people about the reopening economy, Twitter data was collected, representing the 51 states including Washington DC of the US. State-wide socioeconomic characteristics of the people (e.g., education, income, family size, and employment status), built environment data (e.g., population density), and the number of COVID-19 related cases were collected and integrated with Twitter data to perform the analysis. A binary logit model was used to identify the factors that influence people toward a positive or negative sentiment. The results from the logit model demonstrate that family households, people with low education levels, people in the labor force, low-income people, and people with higher house rent are more interested in reopening the economy. In contrast, households with a high number of members and high income are less interested to reopen the economy. The accuracy of the model is good (i.e., the model can correctly classify 56.18\% of the sentiments). The Pearson chi2 test indicates that overall this model has high goodness-of-fit. This study provides a clear indication to the policymakers where to allocate resources and what policy options they can undertake to improve the socioeconomic situations of the people and mitigate the impacts of pandemics in the current situation and as well as in the future. \\

\end{abstract}

\end{frontmatter}

\section{Introduction}\label{Sect:Introduction}
\begin{figure}[htbp]
    \centering
    \includegraphics[width=0.7\linewidth]{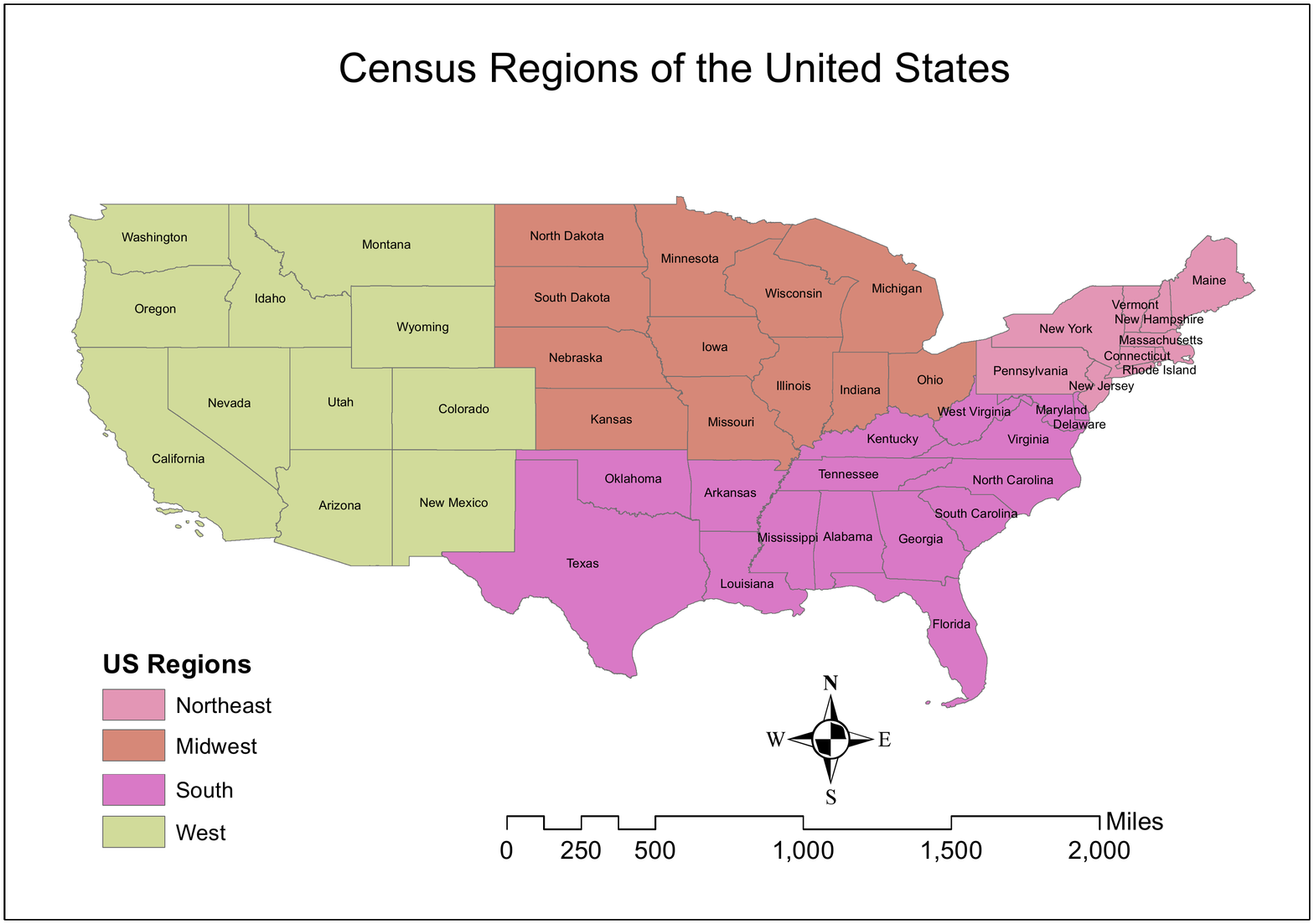}
    \caption{Census regions of the US.}\label{Fig:Regions_US}
\end{figure}

There is a critical need to understand public sentiment concerning post-COVID-19 economic reopening, and the associated socioeconomic factors. Since being first documented in the mid-1960s, there have been seven identified Coronaviruses in the world that can infect humans. Within the human population, Sudden Acute Respiratory Syndrome – Coronavirus-2 (SARS-Cov-2) which causes the disease known as COVID-19 is the fifth endemic Coronavirus including 229E, HKU1, NL63, and OC43 \cite {CDC_new, li2020substantial}. COVID-19 is the highly infectious disease caused by the third identifiable Coronavirus that emerged among humans in the last two decades. Among the three most recent Coronaviruses, the Severe Acute Respiratory Syndrome Coronavirus (SARS-CoV) emerged in China between November 2002 and July 2003 spread in 17 countries with a fatality rate of 9.6\% \cite {munster2020novel, rodriguez2020going}. In 2012, Middle East Respiratory Syndrome Coronavirus (MERS-CoV) was discovered in the Middle East affected 24 countries with a fatality rate of 34.4\% \cite {lai2020assessing, rodriguez2020going}. COVID-19 was first identified in Wuhan, China, by the end of December 2019, already affected over 10 million people in 213 countries of the world with a fatality rate, that had reportedly almost reached 10\% among the closed cases \cite {worldometers_2020}. It is a highly infectious and deadly disease, with widespread transmission and significantly negative impacts on physical, emotional and mental health, economy, and people’s way of living \cite {samuel2020covid, samuel2020feeling}. To control the spread of COVID-19, federal government declared statewide emergency and states governments implemented stay-at-home-order, imposed restriction on mass gathering and non-essential movements. Consequently, people are confined at home with constant fear and uncertainty. Additionally, the unemployment rate showed an increasing trend. To tackle economic depression, many people are arguing to reopen the economy. This study investigates the sentiment of people towards reopening the US economy and finds the underlying socioeconomic factors that are associated with prominent public sentiment. We combine the US public Twitter sentiment and the US demographic information from the US Census Bureau. The studied US Census regions are shown in Fig. \ref{Fig:Regions_US}. 

Continuous lockdown is not a long-term solution for any country. Deliveries of necessary medical supplies including personal Protection Equipment (PPE) and lab equipment are hindered because of the travel restrictions. The supply of necessary foods and household necessities has been halted due to the affected global supply chain system. According to the US Department of Labor, the US economy has lost about 20.5 million jobs with a surge of the unemployment rate to 14.7\% in April 2020 which surpassed the post-World War II record of 10.8\% in 1982 \cite {Lucia_2020}. Labor Body of the United Nations (UN) reported that the world is expected to witness a 6.7\% elimination of working hours, which is equivalent to 195 million jobs, in the second quarter of 2020 due to economic disruption triggered by the COVID-19 pandemic \cite {UN_2020}. This economic downfall describes the depth of the economic recession worldwide caused by COVID-19 and related lockdowns and travel restrictions. Moreover, the persistence of the Coronavirus outbreak is posing a threat to the education systems. Schools, colleges, and universities are closed because of the virus which is hampering in-seat education and hands-on laboratory work severely \cite {bryant2020estimating}. Many local, as well as international students, are the worst victims because they are considered as the potential risk of COVID-19 importation \cite {ma2020travel, igwe2020coronavirus}. However, taking into account the adverse effects of COVID-19, people are craving to go back to normal life and regular activities. The emotionally challenging but true reality is that, amidst the fear of pandemic, people need to go out, do their jobs and run the economy.

Past research investigated fear and trust sentiments of the people towards reopening the economy in the US using exploratory textual analytics, textual data visualization, and hypothesis testing techniques \cite {samuel2020feeling}. Similarly, past research has also investigated and classified sentiments of the users (e.g., positive, negative) towards an item, situation, and system \cite {samuel2020covid}. However, they rarely discussed the underlying socioeconomic factor associations for such sentiments. This study attempts to explore the socioeconomic factor associations for positive and negative sentiments of the people about reopening the economy in the US in the middle of COVID-19 global crisis considering the situational uncertainties (i.e., changes in work and travel pattern due to lockdown policies) and depressions of the people. To control the massive spread of COVID-19 cases by forcing people to stay isolated, the federal and state governments have imposed ‘stay at home’ order and ‘state emergency’. Consequently, dramatic changes have been observed in the daily lifestyle and travel patterns of the people. 

People are adversely affected by the many COVID-19 related anti-transmission measures taken by the state and federal governments. Numerous companies and business are already closed permanently. Many people have lost their jobs and undergoing uncertainty and depression. To address this problem state governments have already started to reopen the economy with different stages of operation (i.e., the complete business operation to the limited scope of operation). However, there is a counter-argument to delay reopening the economy since it will allow people to interact with each other and make different types of trips and consequently the states will face serious difficulties to manage the situations. Many persons are expressing their opinions and aspirations on social media for and against the new normal reopening. This study aimed to understand the sentiment of the people towards the reopening of the economy and investigate the reasons that influence their positive and negative sentiments about the imminent new normal reopening amid COVID-19 pandemic situations.
Our contributions through this paper are as follows: 
\begin{itemize}
    \item Novel data assimilation: We have collected Twitter data to analyze the public sentiment about reopening the US economy amidst COVID-19 pandemic and integrated those Twitter-generated sentiment results with the US Census data \cite{acs_data2019} to understand what socioeconomic factors influence individuals expressing positive or negative sentiments about reopening. 
    \item Early application of methodology: We have provided a detailed methodology of this study from data collection to results discussion with a visual representation. Any potential researchers who wants to collect data from social media (e.g., Twitter, Facebook, LinkedIn, Instagram, and news agency) about a real-world social event (e.g., man-made and natural disasters, political affairs, religious and racial conflicts) and integrate them with Census or household based survey to get some insights, can adopt this study methodology to conduct a similar kind of research.
    \item Parsimonious Logit model: We have modeled a binary logit model with Twitter sentiments and Census data to better understand the most influential features in reopening public sentiment from a set of total 47 initial features. The developed model has over 56\% accuracy to identify the sentiments with a high goodness-of-fit. 
    \item Timely recommendations: Based on our research findings, we have made some suggestions/recommendations for policymakers where to allocate resources to improve the socioeconomic situations of the country and reduce the post COVID-19 sufferings of people.  
\end{itemize}

The rest of the paper is organized as follows: Section \ref{Sect:Literature Review} discusses the literature review of this study. Section \ref{Sect:Data and Study Methods} demonstrates data handling process and modeling binary logit method. Section \ref{Sect:Results} discusses about the results and findings of this study. Finally, we conclude this paper in Section \ref{Sect:Discussions and Conclusions}.

\section{Literature Review}\label{Sect:Literature Review}
The present research seeks to understand factors that support reopening, expressed through positive sentiment towards reopening. The goal of the present study is to extend past research which has demonstrated the prominence of positive sentiment towards reopening, though there exists a fair amount of negative sentiment as well \cite{samuel2020feeling}. Though Twitter data has been intrinsically analyzed extensively and also contextualized to numerous domains, yet past research has not combined recent Twitter data with demographic data to model potential relationships between sentiment classes based on Tweets and COVID-19 relevant data \cite{yu2020prediction,saif2016contextual,sul2017trading}. To provide a meaningful research basis for such an exercise, the present study conducted focused literature review on relevant topics, as summarized in the following sections on Twitter analytics, human behavior and sentiment analysis. 

\subsection{Twitter data analytics}
Twitter data has been used for a wide range of analyses, including but not limited to healthcare, retail marketing, stock trading, education and politics \cite{sinnenberg2017twitter, senti2020feeling, ibrahim2019decoding,kretinin2018going, sul2017trading,wang2019common, samuel2018going, ansari2020analysis}. Twitter data offers a wide range of variables depending on the download programming interface or mechanism used. The use of rsenti package in R allows for the download of 90 variables(including variables such as type of device used, stated location, hast tags, display text width, reply to user ID, quote, retweet, favorite count, retweet count, URLs used, followers count, and date and time, to name a few) providing a rich array of variables associated with each post, which can be used to better understand the sentiment associated with the Tweet \cite{kearney2019rtweet}. In addition to the rich diversity of Twitter variables that lend themselves to analysis, Twitter posts or ``Tweets" contain textual which are not easily manipulated, and therefore requires specialized analysis. Additional elaboration on the specific steps used is provided under the methods section (Section \ref{Subsection:Study_Methods}). 

\subsection{Human behavior and sentiment}
The present study utilizes public sentiment derived from social media posts as a key variable, and this is supported by extant research which has used sentiment analysis for diverse research purposes such as decision support, education, politics, opinion mining, data visualization, healthcare and hate crimes, and the importance of education, gender sensitivity and motivation \cite{samuel2020covid, wang2019common,ansari2020analysis, ravi2015survey,thesis2016analysis, gibbons2019twitter, Conner2019picture, muller2019hashtag, samuel2020beyond, Samuel8Principles}. These studies have used a wide range of methods, tools and languages such as Python and R, and their associated libraries, to estimate sentiment from social media posts. Sentiment estimation can be broadly classified into two buckets, the first is the assignment of a score which ends to be continuous within a given range of an approximate minimum negative value (such as -2)  to an approximate maximum positive value (such as +2), and the second consists of binary classification mechanisms (usually into positive and negative sentiment classes) or categorical classifications of data into sentiment classes such as fear, trust and sadness. For the purposes of this study, we use the R statistical modeling language from CRAN, and its associated sentiment analysis packages called sentimentr and syuzhet \cite{rinker2019package, jockers2017package, r2011r}. Past studies have also developed customized mechanisms to study human characteristic traits such as dominance, with the potential for corresponding emotions of anger and elation expressed through textual communications, and identified via manual or automated textual analytics \cite{liu2012sentiment,samuel2014automating}.

\subsection{US Census data and socioeconomic analysis}
To find out the socioeconomic factor associations of reopening sentiments, this study also collected socioeconomic and demographic (e.g., income, education, age, family type and size, race, housing type etc.) information of the people from the American Community Survey (ACS) \cite {acs_data2019}. Moreover, data on the factor of the built environment (e.g., population density) was collected from ACS to assess the impacts of urban form on the Coronavirus fears and reopening sentiments of the people. ACS is conducted by the US Census Bureau each year to collect vital information about the citizens. The data is free, publicly available, and considered as an important source of information for researchers from different disciplines. Many previous studies collected information from ACS and leveraged with Twitter data to analyze sentiment of the people in the arena of public health \cite {gibbons2019twitter}, urban spaces \cite {oliveira2020outdoorsent}, politics \cite {bail2018using, shor2019political}, disasters management \cite {reynard2019harnessing}, racial conflicts \cite {muller2019hashtag, nguyen2018twitter}, and gender disparity \cite {shor2019women}. Thus, linking Twitter data with Coronavirus data is a common practice among the researchers to evaluate the impacts of socioeconomic and demographic characteristics on the sentiments of the people towards a subject of interest. Moreover, the name of the four regions from where the tweets were generated was collected based on the US Census to evaluate regional impacts on reopening sentiments \cite {regions_data2010}. Dummies (0 or 1) of the regions were created to include them in the model. This study also collected information on the number of cases and deaths from Worldometer \cite {worldometer_2020} to understand how the severity of Coronavirus influence the sentiment of the people about reopening the economy. Considering the unavailability of the exact location of Twitter users, state averaged Census and Coronavirus data were collected and integrated with Twitter data to perform the analysis. The process of Twitter, Census and Coronavirus data integration has been illustrated in Section \ref{Sect:Data and Study Methods}. Table \ref{Table:Descriptive} represents the descriptive statistics of the dependent and independent variables used in the binary model.

\subsection{Reopening risk assessments}

Restarting the US businesses would cause a rise in the mortality \cite {dyer2020covid}. A study using agent-based Susceptible, Exposed, Infected, and Recovered (SEIR) model \cite {killian2020evaluating} on the population-specific data (such as contact patterns, household structure, age distribution, and comorbidity rates) evaluated an alternative lockdown and reopening scenarios for three specific states in the US (Florida, Georgia, and Mississippi). The model assessed that imposing lockdown one week earlier in all the three states could have saved hundreds of lives from COVID-19. However, to reopen the economy even with a limited capacity it is required to reduce the population contact down to 20-25\% and implement strict social distancing measures along with the use of personal protective equipment \cite {killian2020evaluating}. Besides, a robust testing capacity would help the policymakers to estimate more precise reopening dates and it would be beneficial to detect and isolate asymptomatic carriers in a quicker manner after reopening. Various empirical evidence, trials, and observations suggested that the proper use of medical masks, combined with other non-pharmaceutical interventions (NPIs) such as thorough handwashing and strict social distancing, testing, contact tracing, and quarantine, is an effective way to reduce or interrupt the transmission of respiratory infections of COVID-19. Therefore, the deployment of masks in the public zones combined with other measures may eventually help in reopening the economy and transitioning into the post COVID-19 world \cite {polyakovacan2020, yamana2020projection}. 

\begin{table}[htbp]
\small
\centering
\caption{Descriptive statistics of the variables.}\label{Table:Descriptive}
\begin{tabular}{lp{4cm}lllll}
\toprule
\textbf{Variable} & \textbf{Variable description}                                       & \textbf{Measure} & \textbf{Mean} & \textbf{SD} & \textbf{Min} & \textbf{Max} \\ \toprule
\multicolumn{7}{l}{Tweet characteristics (2501   Tweets)}                                                                                                              \\ \midrule
Sentiment         & Sentiment type                                                      & Dummy (0, 1)     & 0.48          & 0.50        & 0.00         & 1.00         \\ \hline
TW                & Number of words in the tweet                                        & \#               & 169.35        & 81.49       & 6.00         & 296.00       \\ \midrule
\multicolumn{7}{l}{Regional Dummies (4 regions)}                                                                                                                       \\ \midrule
NE                & Northeast regions                                                   & Dummy (0, 1)     & 0.20          & 0.40        & 0.00         & 1.00         \\ \hline
MW                & Midwest region                                                      & Dummy (0, 1)     & 0.17          & 0.37        & 0.00         & 1.00         \\ \hline
WEST              & West region                                                         & Dummy (0, 1)     & 0.26          & 0.44        & 0.00         & 1.00         \\ \midrule
\multicolumn{7}{l}{State-level socioeconomic   characteristics and population density (51 States including Washington DC)}                                             \\ \midrule
L\_FHH            & Log of Percentage of family   household                             & \%               & 4.18          & 0.05        & 3.77         & 4.32         \\ \hline
AFS               & Average family size                                                 & \#               & 3.26          & 0.17        & 2.85         & 3.62         \\ \hline
EDU2              & Percentage of persons   with high school graduate and some college  & \%               & 47.10         & 4.73        & 30.02        & 59.14        \\ \hline
EDU3              & Percentage of persons with an associate   degree                    & \%               & 8.25          & 1.19        & 3.01         & 11.48        \\ \hline
AGE2              & Percentage of person Age under 18                                   & \%               & 22.25         & 1.66        & 18.10        & 29.50        \\ \hline
WP                & Percentage of white persons                                         & \%               & 75.48         & 8.39        & 25.60        & 94.60        \\ \hline
OCH               & Percentage of owner-occupied housing                                & \%               & 62.66         & 5.77        & 41.80        & 72.90        \\ \hline
PWHI              & Percentage of persons under age   65 years without health insurance & \%               & 10.22         & 4.35        & 3.20         & 20.00        \\ \hline
LF                & Percentage of the population age   16 years+ in the labor force     & \%               & 62.94         & 2.73        & 53.10        & 69.70        \\ \hline
L\_POPDEN         & Log of Population density                                           & Persons/mile2    & 5.24          & 0.99        & 0.18         & 9.20         \\ \hline
CASES             & Total number of cases per 1   million population                    & cases/1M people  & 5856.56       & 5466.00     & 458.00       & 19479.00     \\ \hline
PR                & Persons in poverty (Poverty rate)                                   & \%               & 13.07         & 2.06        & 7.60         & 19.70        \\ \hline
MHHI              & Median household income   (2014-2018)                               & \$               & 61962.71      & 10709.20    & 48.49        & 82604.00     \\ \hline
GR                & Median gross rent (2014-2018)                                       & \$               & 1084.67       & 217.14      & 711.00       & 1566.00    \\ \bottomrule 
\end{tabular}
\end{table}

\section{Data and Study Methods}\label{Sect:Data and Study Methods}
\subsection{Data}
\begin{figure}[htbp]
    \centering
    \subfloat[Tweet Count.]{\includegraphics[width=0.5\linewidth]{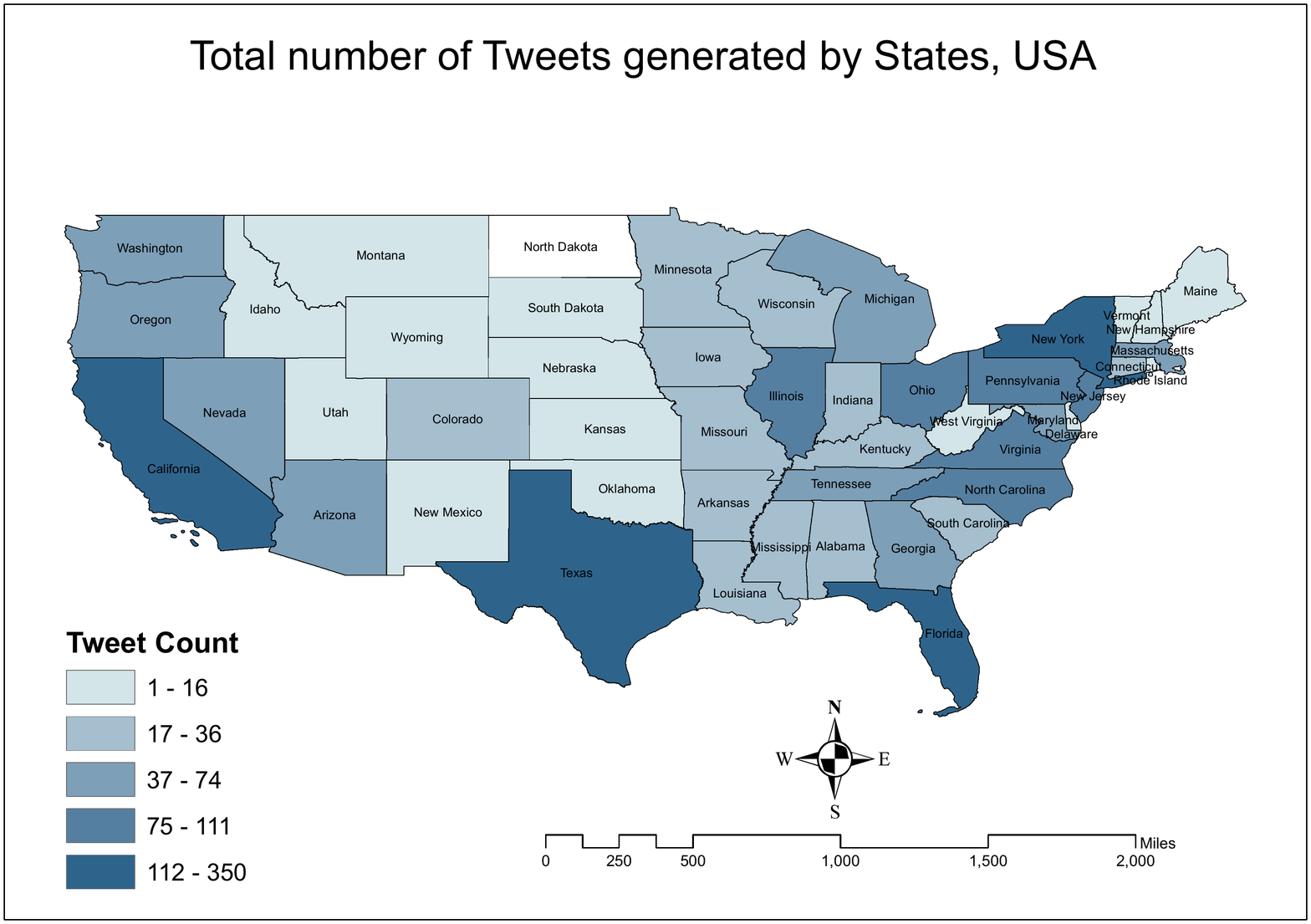}\label{Fig:Tweet Count}}
    \subfloat[Sentiment Score.]{\includegraphics[width=0.5\linewidth]{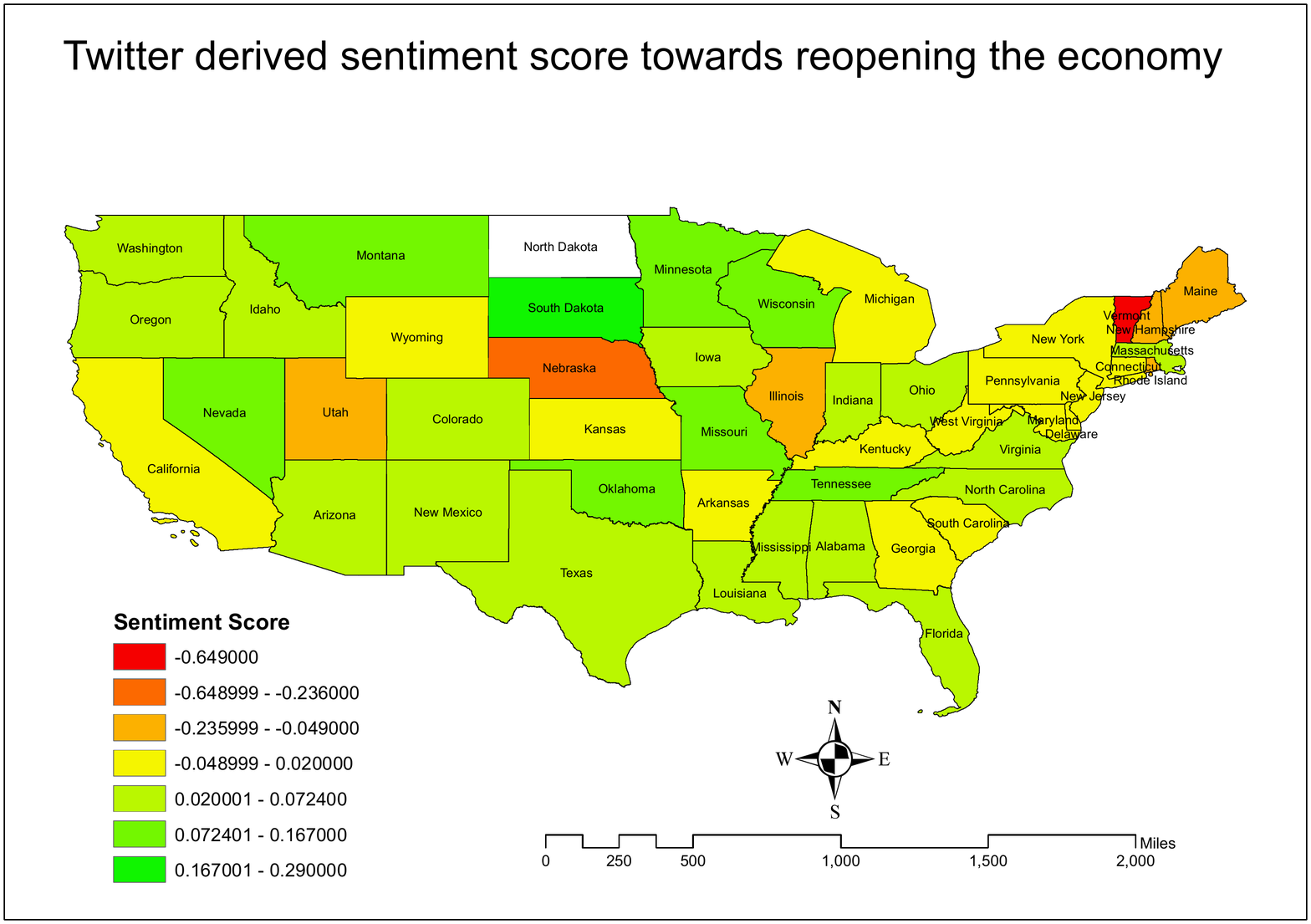}\label{Fig:Sentiment Score}}
    \caption{Number of tweets and Twitter-driven sentiment score.}
\end{figure}

This study uses Twitter data collected between April 30, 2020, and May 08, 2020, to understand the sentiment of the people towards the reopening of the US economy \cite {samuel2020feeling}. A total of 293,597 tweets with 90 variables were downloaded using the keyword “reopen”. The detailed methodology of data acquisition, saving, cleaning, and filtering for obtaining a subset of them that were generated from different states of the US has been described in Section \ref{Subsection:Study_Methods}. After systematic cleaning and filtering, a final dataset consisting of 2507 tweets and twenty-nine variables were used for sentiment analysis and exploring the socioeconomic factor associations that influence sentiments of the people towards reopening the economy. Fig. \ref{Fig:Tweet Count} shows total number of tweets collected from each states in the US. The figure demonstrates that a large number of tweets were generated from the Western and Northeastern regions of the US. Most of the tweets were collected from California, Texas, New work, Florida, Pennsylvania, Illinois, Ohio, North Carolina, Virginia, New Jersey, Arizona, and Nevada. In contrast, no tweets were collected from North Dakota. Fig. \ref{Fig:Regions_US} displays the Census regions of the US with the states in each region.

Sentiment analysis is a popular topic of research among the researchers after collecting data from Twitter, webpage, product reviews, newspaper etc. \cite {ravi2015survey, severyn2015twitter, liu2012emoticon, go2009twitter}. The primary objective of the sentiment analysis is to investigate opinions, attitudes, and emotions of the users (i.e., positive or negative feeling) towards a subject matter of interest (e.g., entity, person, issue, event, and topic). Sentiment analysis is beneficial for both customers and service providers to improve the mutual relationships \cite {ravi2015survey}. In this study, Twitter data was analyzed to understand the sentiments of the Americans towards the new normal reopening amidst COVID-19 outbreak. Using R packages Syuzhet and sentimentr, sentiments were classified and assigned a score based on matching keywords, word sequences, and prewritten lexicons. Sentiment score was assigned within a range of -2 to +2. The maximum negative value indicates negative sentiment, whereas the maximum positive value indicates positive sentiment, with a score 0 indicates neutral sentiment. Results from a preliminary analysis showed that about 48.27\% of the users expressed positive sentiment. In contrast, about 36.82\% and 14.92\% of the users expressed negative and neutral sentiments, respectively. Fig. \ref{Fig:Sentiment Score} shows spatial distribution of the Twitter-driven sentiment score towards reopening the economy. It indicates that most of the states showed positive sentiment towards reopening of the economy. Calculating an average score value for the US, we found that mean value of the sentiment score is 0.0271 considering positive, negative and neutral tweets. Thus, most of the Twitter users posted positive information about the reopening. States with highest positive sentiments include South Dakota, Wisconsin, Oklahoma, Montana, Tennessee, Minnesota, and Missouri. On the other hand, Vermont, Nebraska, Utah, Maine, Illinois, New Hampshire, and Rhode Island showed the highest negative sentiment towards the reopening.

\subsection{Study methods}\label{Subsection:Study_Methods}
\begin{figure}[htbp]
    \centering
    \includegraphics[width=0.57\linewidth]{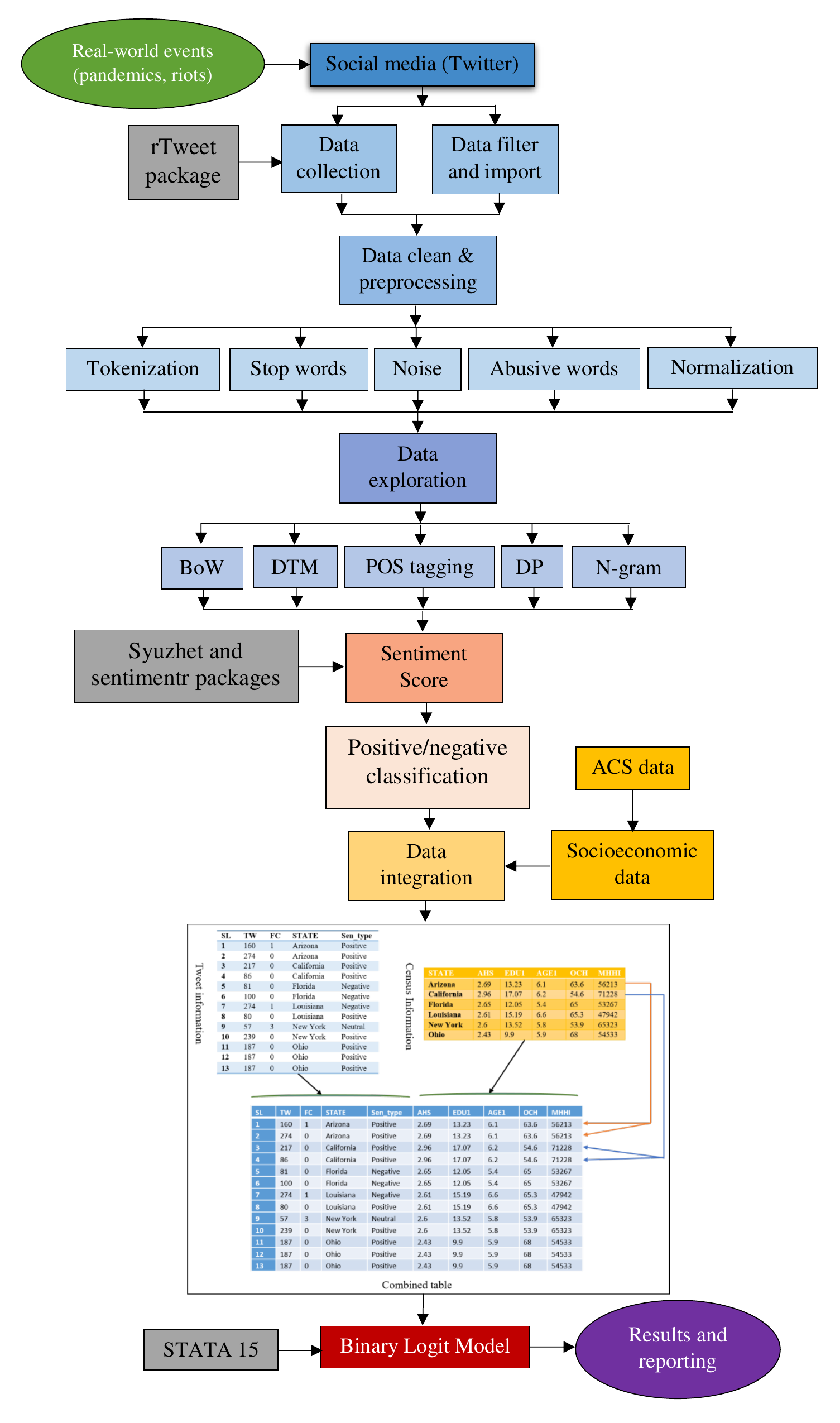}
    \caption{Study design flowchart.}\label{Fig: Flow_Chart}
\end{figure}

A detailed description of the systematic methodologies adopted in this study, starting from data collection 
to results discussion and reporting, is shown in Fig. \ref {Fig: Flow_Chart}. We collected data from Twitter to understand feelings of the people by using the rTweet package in R and associated Twitter API. However, this method is generally applicable for collecting data from any social media platforms (e.g., Facebook, LinkedIn, Instagram, and news agency) regarding any real-world social events (e.g., manmade and natural disasters, political affairs, religious and racial conflicts). After filtering the tweet's information, the data is saved in CSV format for subsequent prepossessing. Data preprocessing (e.g., cleaning, removing noises) is an important step in text analysis and classification. In the data preprocessing steps, unnecessary words, such as stop words (i.e., pronouns, articles, prepositions such as ‘the’, ‘a’, ‘about’, ‘we’, ‘our’, etc. that do not have any significant contribution to text classification), noises (i.e., punctuation, special characters), and abusive words (i.e., slang words) were removed from the text to improve the efficiency of the system \cite {kowsari2019text}, \cite {welbers2017text}.

Moreover, tokenization and normalization techniques are applied to process the text. Tokenization is the process of dividing texts into words, phrases, and symbols which are known as tokens \cite {verma2014tokenization}. The main goal of the tokenization is to find out the words or group of words in a sentence, which is the foundation of text analysis. Tokenization is very important for text analysis because a meaningful computation primarily depends on the components (tokens) of the text, not on the full text. For example, if there is a sentence like ‘we have collected data from twitter’. After tokenization, the set of tokens will be: \{‘we’, ‘have’, ‘collected’, ‘data’, ‘from’, ‘twitter’\}. Normalization transforms words into a common form that allows the computer to identify the same words with similar meaning and remove one of them \cite {welbers2017text}, consequently significantly reduces data size and increases computational efficiency. Lowercasing, stemming and lemmatization are three important and useful methods of text normalization. Lowercasing converts every letter to lowercase and make similar words consistent to each other \cite {kowsari2019text}. For example, lowercasing converts each letter of ‘An’ and ‘an’ to lowercase and helps the computer to identify the two words are identical. Stemming converts words to their base form (stem) \cite {singh2016text}. The same word appears in different forms in the texts that are consolidated into the same features \cite {mawardi2018fast}. For example, the semantic meaning of ‘read’ and ‘reading’ is the same, thus they are combined as ‘read’ to avoid any confusion. Lemmatization is the most advanced method of text normalization, which replaces words with their same morphological root using a dictionary \cite {welbers2017text}. Lemmatization is very similar to stemming. However, the basic difference between them is that lemmatization does not produce a stem but replace the suffix of the word to normalize it with usually a different word suffix \cite {plisson2004rule}. Normalized words with both stemming and lemmatization could be the same. For example, normalized word of ‘reads’, ‘reads’, and ‘reading’ is the same in both methods. However, normalization of the same words could be different in stemming and lemmatization. For instance, the normalized word of computes, computing, computed is ‘comput’ in stemming and compute in lemmatization. After processing the data, different data exploration techniques 
were used to extract insights from the Twitter data as stated below: 

\begin{enumerate}
    \item \textbf{Bag-of-words (BoW)} count different word frequencies in the text to determine the focal point of the text analysis, ignoring the order of the words \cite {kowsari2019text}. For example, the sentence “Although the order of the words is ignored, multiplicity is counted and used to determine the focal point of the text analysis” can be tokenized as \{‘Although’, ‘the’, ‘order’, ‘of’, ‘the’, ‘words’, ‘is’, ‘ignored’, ‘multiplicity’, ‘is’, ‘counted’, ‘and’, ‘used’, ‘to’, ‘determine’, ‘the’, ‘focal’, ‘point’, ‘of’, ‘the’, ‘text’, ‘analysis’\}. Thus, the corresponding BoW is \{1, 4, 1, 2, 1, 2, 1, 1, 1, 1, 1, 1, 1, 1, 1, 1, 1\}.
    \item \textbf{Document term matrix (DTM)} is used to represent text corpus (i.e., collection of texts) in a bag-of-words \cite {welbers2017text}. DTM is a matrix where rows contain documents, columns contain terms, and cells contain the number of each term that occurred in each text document. TDM allows the researchers to analyze data with vector and matrix algebra, which effectively convert text to numbers. Moreover, text data stored in the DTM format improve memory efficiency and optimize the operation of the data analysis.
    \item \textbf{Parts-of-speech (POS)} tagging is a basic part of the syntactic analysis, which has numerous application in NLP \cite {gimpel2011part}. In POS tagging, worlds in the text such as nouns, verbs, articles, and adjectives are identified to understand the context of the text \cite {welbers2017text}. For example, tagging nouns and proper names researchers identify similar events in news items. It is also a good approach to remove articles and pronouns from the text, which has no meaningful role in the text analysis.
    \item \textbf{Dependency parsing (DP)} illustrates the syntactic relationship between different tokens \cite {van2008semantic}. For example: ‘John is an assistant professor at UNCC and he is very popular among the students’. In the example, it is understood that ‘John’ is a nominal subject and ‘professor’ is an adjective, which indicates that John is a professor who teaches students.
    \item \textbf{The N-gram} technique is used to understand the association between the words. N-gram is not a representation of text, rather presents a set of n-words in the text with their order \cite {kowsari2019text}. N-gram techniques tokenize texts into single words (unigrams), sequences of two words (bigrams), three words (trigrams), and so on to maintain the order of words and syntactical properties \cite {welbers2017text}.

\end{enumerate}
The findings of the exploratory analysis using steps and processes described above helped to gain a clearer understanding of public perspectives on reopening \cite{samuel2020feeling}. After data exploration, sentiment score was generated for each tweet by using the R package sentimentr and sentiments were classified into positive, negative, and neutral based on matching keywords, word sequences, and prewritten lexicons. Twitter data was then integrated with state-wide averaged socioeconomic data collected from the Census to conduct the analysis. A binary logit model was used to evaluate the factors that influence people's sentiment towards a new normal reopening. To perform the analysis, the categorical variable of tweet sentiments was converted to dummy variables where positive sentiment was assigned a value of ‘1’ and negative and neutral sentiments were assigned a value of ‘0’. The category of the tweets was used as the dependent variable in the binary model. On the other hand, different socioeconomic and demographic variables, regional dummies, and Coronavirus cases and deaths have been used as the independent variables in the binary logit model. Finaly, the model results are discussed and reported to get some insights about the reopening sentiment of the people. The following subsection discusses about the theoretical framework of binary logit model.

\subsection{Binary logit model}
In the linear regression model, the response variable Y is quantitative. However, in many situations, the response variable is rather qualitative or categorical, for instance, the sentiment of a tweet could be categorized into positive or negative. The logistic regression model also known as \emph{logit model} classifies the sentiment based on the probability. Assume $X$ is the set of features, $[x_1,x_2,\cdots,x_n]$,  where $n$ is the total number of features in reopening sentiment analysis. $pr(Y=1|x)$ denotes the probability of positive sentiment about reopening given the feature $x$. Conversely, $pr(Y=0|x)$ denotes the probability of negative reopening sentiment given the feature $x$. Using the linear regression model, $pr(Y=1|X)=P(X)$ can be computed as,
\begin{equation}\label{Eq:Linear_regression}
    P(X)=\beta_0+\beta_1X
\end{equation}
where $\beta_0$ is the intercept and $\beta_1$ is the co-efficient of $X$. However, Eq. (\ref{Eq:Linear_regression}) may predict $P(X)<0$ for $X$ close to $0$ and $P(X)>1$ for large value of $X$. To bound P(X) in the range 0 and 1, we use the following logistic function \cite{ILSR2017}, 
\begin{equation}\label{Eq:Logistic_function}
    P(X)=\frac{e^{\beta_0+\beta_1X}}{1+e^{\beta_0+\beta_1X}}
\end{equation}
Consider the boundary value is 0.5. We get the following estimated probability ($\hat{P}(X)$) from Eq. (\ref{Eq:Logistic_function}), 
\begin{equation}\label{eq:estimated_prob}
    \hat{P}(X)=\begin{cases}
    1 & \text{if $P(Y=1|X)\geq 0.5$}\\
    0 & \text{otherwise}
    \end{cases}
\end{equation}
After some mathematical manipulation from Eq. (\ref{Eq:Logistic_function}), we can get, 
\begin{equation}\label{Eq:odds}
    \frac{P(X)}{1-P(X)}=e^{\beta_0+\beta_1X}
\end{equation}
After taking $\log$ on both side of Eq. (\ref{Eq:odds}) we get, 
\begin{equation}\label{Eq:log_odds}
    \log\left(\frac{P(X)}{1-P(X)}\right)=\beta_0+\beta_1X
\end{equation}
The left-hand side of Eq. (\ref{Eq:log_odds}) is called log-odds or logit, which shows that logistic regression model has a logit which is linear in $X$. The coefficients $\beta_0$ and $\beta_1$ are estimated using the maximum likelihood method. The idea is to estimate the value of $\beta_0$ and $\beta_1$ for each feature ($x_i$) so that it minimizes the difference between the predicted probability $\hat{P}(x_i)$ and observed probability $P(x_i)$. The likelihood function can be expressed as follows \cite{ILSR2017},
\begin{equation}\label{Eq:Likelihood_function}
    L(\beta_0,\beta_1)=\max \left( \prod_{i:y_i=1}P(x_i).\prod_{i^\prime:y_{i^\prime}=0}\left(1-P(x_{i^\prime})\right)\right)
\end{equation}
To increase the computation speed, we take $\log$ on Eq. (\ref{Eq:Likelihood_function}) and get the following log likelihood function,
\begin{equation}\label{Eq:Log_Likelihood_function}
    LL(\beta_0,\beta_1)=\max \left( \log\sum_{i:y_i=1}P(x_i)+\log\sum_{i^\prime:y_{i^\prime}=0}\left(1-P(x_{i^\prime})\right)\right)
\end{equation}
In other words, estimate the values of $\beta_0$ and $\beta_1$ so that Eq. (\ref{Eq:Log_Likelihood_function}) yields the maximum value. 

\section{Results}\label{Sect:Results}
\subsection{Calibrated model}

The binary logit model is calibrated and estimated using STATA 15 software \cite {Stata_15}. The log-likelihood method is used to calculate the coefficients. The final results of the model indicating the impacts of independent variables on the dependent variables are presented in Table \ref{Table2:Results_logit}. The table also reports the standard error, z-value, and probability level (P-value) of the estimates. Many of the coefficients are statistically significant at 0.001, 0.05, and 0.10 levels. However, some of the coefficients with a P-value greater than 0.10 are retained in the model to portray the relationships between dependent and some statistically insignificant independent variables, yet important factors that can influence sentiment of the persons.

\begin{table}[htbp]
\centering
\caption{Results of the binary logit model.}\label{Table2:Results_logit}
\begin{tabular}{lllll}
\toprule
\textbf{Sentiment}     & \textbf{Coef.} & \textbf{Std. Err.} & \textbf{z} & \textbf{P\textgreater{}z} \\
\midrule
TW                     & 0.003          & 0.001              & 5.990      & 0.000                     \\ 
NE                     & 0.253          & 0.246              & 1.030      & 0.303                     \\ 
MW                     & 0.250          & 0.221              & 1.130      & 0.257                     \\ 
WEST                   & 0.702          & 0.282              & 2.480      & 0.013                     \\ 
L\_FHH                 & 2.414          & 2.013              & 1.200      & 0.230                     \\ 
AFS                    & -3.331         & 1.021              & -3.260     & 0.001                     \\ 
EDU2                   & 0.034          & 0.026              & 1.310      & 0.189                     \\ 
EDU3                   & 0.062          & 0.054              & 1.150      & 0.249                     \\ 
AGE2                   & 0.006          & 0.099              & 0.060      & 0.953                     \\ 
WP                     & -0.006         & 0.008              & -0.820     & 0.415                     \\ 
OCH                    & 0.011          & 0.022              & 0.520      & 0.600                     \\ 
PWHI                   & 0.057          & 0.024              & 2.360      & 0.018                     \\ 
LF                     & 0.157          & 0.058              & 2.700      & 0.007                     \\ 
L\_POPDEN              & 0.122          & 0.098              & 1.240      & 0.213                     \\ 
CASES                  & -0.000003      & 0.000              & -0.180     & 0.860                     \\ 
PR                     & 0.206          & 0.081              & 2.530      & 0.011                     \\ 
MHHI                   & -0.00002       & 0.000              & -2.130     & 0.033                     \\ 
GR                     & 0.003          & 0.001              & 2.230      & 0.026                     \\ 
Constant               & -18.042        & 10.131             & -1.780     & 0.075                     \\ 
LR chi2(18)            & \multicolumn{4}{l}{63.390}                                                   \\ 
Prob \textgreater chi2 & \multicolumn{4}{l}{0.000}                                                    \\ 
Pseudo R2              & \multicolumn{4}{l}{0.018}                                                    \\ 
Log-likelihood         & \multicolumn{4}{l}{-1700.284}  \\ \bottomrule   
\end{tabular}
\end{table}

\subsection{Discussion of results}    
Results presented in Table \ref{Table2:Results_logit} show that longer tweets (0.003) are more likely to indicate positive sentiment about the reopening. The influence of tweet width on positive sentiment is statistically significant at P-value of 0.05. Thus, people are more likely to post a longer statement on Twitter to express their positive feelings about the reopening of the economy with useful information. A study analyzed Twitter data to understand sentiments of the citizens for allocating resources during Hurricane Irma in 2017 in Florida \cite {reynard2019harnessing}. Upon analyzing data the study found that longer tweets are more likely to have useful information with sentiment contents. It also revealed that popular tweeters were more likely to have positive sentiments and less likely to have useful information about the disaster. Thus, longer tweets can provide insightful information on the public sentiments and can be used for crises management during any man-made and natural pandemics.

\subsubsection{Regional, family and education association}
People live in the Northeast (0.253) and Midwest (0.250) regions are more likely to express positive sentiment about the reopening. However, the relationship is not statistically significant at P-value of 0.05. Thus, tweets generated from the Northern and Midwest regions have less sentiment contents. Rather, the states located in these two regions, particularly, New York, New Jersey, Illinois, Massachusetts, Pennsylvania, and Michigan are more concerned about the health condition of family members and relatives, and the COVID-19 pandemic due to higher number of Coronavirus cases and deaths. People live in the West region (0.702) are more likely to stay positive to reopening the economy and this relationship is statistically significant at P-value of 0.05. The people in the western regions mainly live in California, Nevada, Oregon, and Washington are more interested to reopen the economy because of higher monthly house rent, a higher number of foreign-born people, and low population density than other regions.

Family households (2.414) are more likely to express positive sentiment about the reopening compared to non-family households. However, the relationship is statistically significant at a marginal level. Most of the family households want reopening of the workplaces to earn family expenses. However, they are also concerned about the COVID-19 related health risks associated with the reopening of the workplaces. People with low education levels, high school graduates, and some college (0.034) and associate degree (0.062), are more likely to express positive sentiment about the reopening of the economy. People with lower levels of education are the worst victim of COVID-19. Most of them have lost their jobs due to the closure of workplaces. Moreover, they have limited options to work from home because most of these low paid jobs are on-site in nature. However, both of them are statistically significant at a marginal level.

\subsubsection{Age and income}
Young people (age under 18 years) (0.006) is positively associated with the reopening of the economy. Thus, younger generations are more likely to see a new normal reopening where they can enjoy a COVID-19-restriction-free life. However, the relationship is not statistically significant at P-value of 0.05. Thus, tweets posted by younger people are less likely to have any sentiment contents related to reopening. However, previous study recommended to take into consideration of over-reporting tendency of the young generation to gauge the real sentiment of the pandemic \cite {reynard2019harnessing}. Persons age 16 years and above involved in the labor force (0.157) are more likely to reopen the economy. The relationship is statistically significant at P-value of 0.05. Similarly, persons under 65 years without health insurance (0.057) are more positive to reopen the economy. The relationship is statistically significant at P-value of 0.05. A previous study mentioned that elderly persons having fragile health conditions are more prone to the risk of severe illness from COVID-19 than other age cohorts \cite {ma2020travel}. Thus, working-age people usually age under 65 years are positive about the reopening despite not having health insurance and tweets posted by them often contain positive facts and figures in favor of a pre-COVID-19 scenario.

Low-income people (0.206) are positively associated with reopening the economy with a statistically significant P-value of 0.05. Thus, people with low household income are more interested in reopening the economy.
Similarly, people with high median gross household rent (0.003) are more likely to post tweets with positive sentiment concerning reopening the economy which is statistically significant at P-value of 0.05.
In contrast, people with high median household income (-0.00002) are less likely to reopen the economy. The relationship is statistically significant at P-value of 0.05. The people of this stratum mostly work in the Information Technology (IT) sector and at the high positions with corporate organizations. Usually, they are well paid and have an option to continue work from home. Thus, they are less affected by the COVID-19 compared to the low-income people and consequently less interested or apathetic to reopen the economy. The results of this study conform to previous study \cite {reynard2019harnessing} which reported that higher per capita income increases the probability of the tweets to have negative sentiment about a pandemic.

\subsubsection{Family size and other factors}
Average family size (-3.330) is negatively associated with the reopening of the economy which indicates that families with a large number of members are less likely to reopen the economy. The relationship is statistically significant at P-value of 0.05. There is a high possibility of COVID-19 infection among the households with larger members compared to signal age group (i.e., single household) \cite {wilder2020role}, thus they are less interested to reopen the economy considering the rapid transmission of disease through social contact. Similarly, white people (-0.006) are negatively associated with the reopening. Thus, white people are less likely to express their opinion in favor of reopening because they are in advantageous position than their counterparts. However, the relationship is not statistically significant at P-value of 0.05 which indicates that tweets generated by them are less likely to have any sentiment contents. The number of Coronavirus cases (-0.000003) is negatively associated with the reopening. Thus, with an increasing number of cases, people are less likely to reopen the economy. However, the relationship is statistically insignificant at the P-value of 0.05. Moreover, the tweets generated from the states with a higher number of Coronavirus cases and deaths are less likely to have any reopening sentiments. Rather, they are more concerned about the risks associated with COVID-19 and tweets about the severity of the pandemic.

\subsubsection{Fit}
The overall fit of the calibrated model is evaluated based on several key goodness-of-fit statistics. The likelihood ratio chi-square (LR chi2) statistic of the estimated model is 63.39 (Table \ref{Table2:Results_logit}). A lower value of the Chi-square indicates a better fit of the model. P-value (\textless0.000) of the Chi-square statistic indicates that overall the model is statistically significant and better than a model with no predictors. Thus, the model can significantly fit the observed data because the P-value is less than 0.000. 

\begin{table}[htbp]
    \centering
    \subfloat[Pearson chi-square fit statistics.]{
    \begin{tabular}{@{}ll@{}}
\toprule
Goodness-of-fit   test         & Fit   statistics \\ \midrule
Number   of observations       & 2501             \\
Number   of covariate patterns & 1968             \\
Pearson   chi2(1949)           & 2002.96          \\
Prob   \textgreater chi2       & 0.1929  \\
\bottomrule
\end{tabular}\label{Table:Chi_square}
}
\hspace{1 cm}
    \subfloat[Classification summary.]{
    \begin{tabular}{@{}ll@{}}
\toprule
Measure & Outcome \\
\midrule
Correctly classified & 56.18\% \\
Sensitivity          & 49.59\% \\
Specificity          & 62.32\% \\
\bottomrule
\end{tabular}
    \label{Table:Classification}
    }
    \caption{Fit statistics and classification summary.}\label{Table:Classification_Stat}
\end{table}

Other fit statistics also confirm the overall fit of the estimated model (Table \ref{Table:Chi_square}). We calculated Pearson chi-square fit statistics to evaluate the overall fit of the model. This is the formal test of the null hypothesis to assess whether the fitted model is correct. The P-value of this hypothesis testing ranges between 0 and 1. P-value specified $\alpha$ level (i.e., 0.05) indicates that the model is not statistically significant and acceptable. A higher P-value indicates a better fit of the model. Pearson chi-square test statistics (Prob \textgreater chi2 = 0.1929) presented in Table \ref{Table:Chi_square} indicates that we cannot reject the null hypothesis and hence, the model is overall fit. In a nutshell, the above discussed fit statistics indicate that the model can adequately fit the observed data.

We also reported classification statistics to evaluate the accuracy and efficiency of the model (Table \ref{Table:Classification}).  Overall, the model can correctly classify 56.18\% of the sentiments with 62.32\% of specificity (i.e., correctly classify positive sentiment) and 49.59\% sensitivity (i.e., correctly classify negative sentiment). Thus, the classification statistics indicate a good prediction accuracy of the model.

\begin{figure}[htbp]
    \centering
    \subfloat[Normal QQ plot.]{\includegraphics[width=0.45\linewidth]{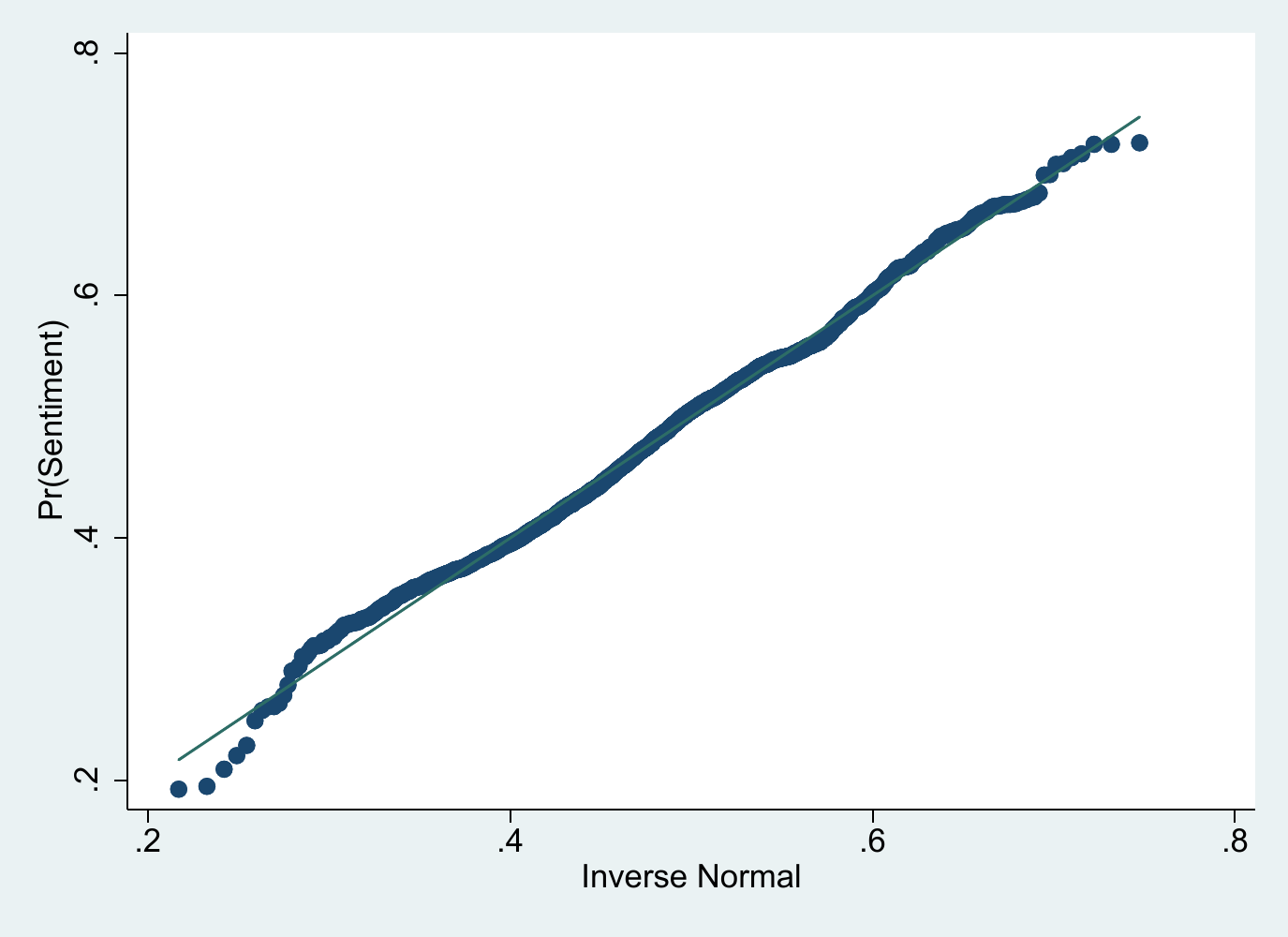}\label{Fig:Normal_QQ_Plot}}
    \hspace{0.7cm}
    \subfloat[Normal probability plot.]{\includegraphics[width=0.45\linewidth]{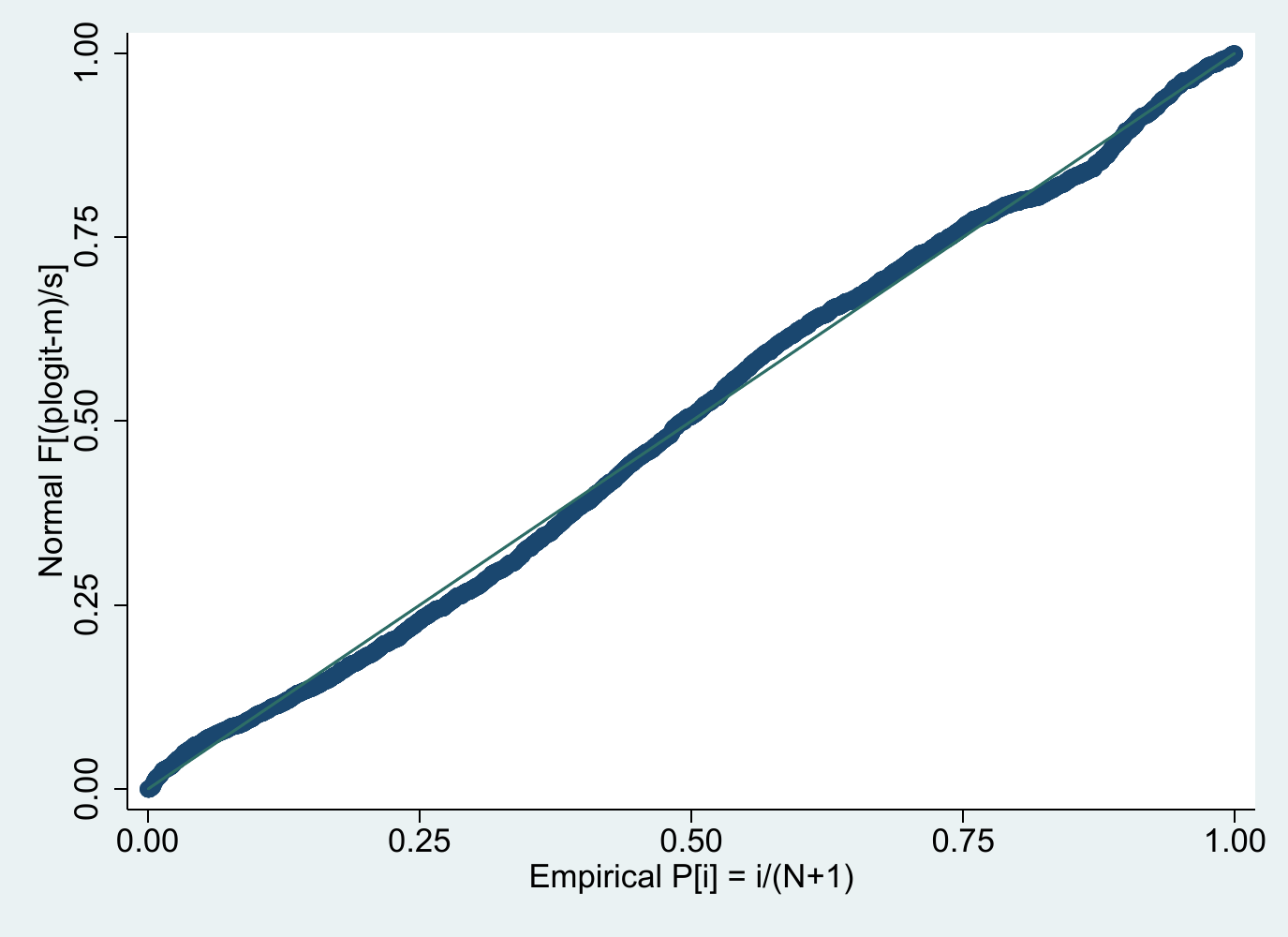}\label{Fig:Normal_probability_plot}}
    \caption{Normal distribution of the residuals.}\label{Fig:Normal_Distribution}
\end{figure}

Normal distribution of the residuals is shown in Fig. \ref{Fig:Normal_Distribution}. The Normal QQ plot and normal probability plot show that the residuals fall perfectly along a linear line at $45^0$ angle. Thus, the residuals are normally distributed. Normal distribution of the residuals indicates that the amount of error in the model is consistent across the observed dataset. Therefore, the predictive capability of the explanatory variables is same for the full range of dependent variables. Moreover, the model can explain all the variations in the dataset sufficiently.

\subsection{Marginal effects}
The results of the binary logit model presented in Table \ref{Table2:Results_logit} provide an indication of the effects (i.e., positive or negative) of independent variables on the dependent variable. However, it is difficult to interpret the magnitude of marginal effects. Hence, we calculated the marginal effects of explanatory variables using the ‘margins’ command in Stata. Margins make result interpretation easier and report elasticities (i.e., percentage change in the likelihood of positive sentiment for a unit change in an explanatory variable). The results of the marginal effects of explanatory variables have been presented in Table \ref{Table:Marginal_effect}. The results in Table \ref{Table:Marginal_effect} indicate that
a 1\% increase in tweet generation from the west region of the US increases the probability of positive sentiment for the reopening by 0.175\%. The results also indicate that employment status has positive impacts on the reopening. 
People with low household income are associated with the positive reopening sentiment. A 1\% increase in low-income people increases the probability of positive sentiment for reopening by 0.051\%. Similarly, high house rent motivates people toward the positive sentiment of reopening the US economy. The result indicates that a 1\% increase in house rent leads to the probability of a 0.001\% increase in positive sentiment.
Interestingly, increasing Coronavirus cases have a limited affect on reopening sentiment - A 1\% increase in Coronavirus cases reduces positive sentiment by only 0.000001\%. 
\begin{table}[htbp]
\centering
\caption{Marginal effect of the explanatory variables.}\label{Table:Marginal_effect}
\begin{tabular}{@{}lllll@{}}
\toprule
Sentiment & dy/dx     & Std. Err. & z      & P\textgreater{}z \\
\midrule
TW        & 0.001     & 0.0001    & 5.990  & 0.000            \\
NE        & 0.063     & 0.061     & 1.030  & 0.303            \\
MW        & 0.062     & 0.055     & 1.130  & 0.257            \\
WEST      & 0.175     & 0.071     & 2.480  & 0.013            \\
L\_FHH    & 0.603     & 0.503     & 1.200  & 0.230            \\
AFS       & -0.832    & 0.255     & -3.260 & 0.001            \\
EDU2      & 0.008     & 0.006     & 1.310  & 0.189            \\
EDU3      & 0.016     & 0.014     & 1.150  & 0.249            \\
AGE2      & 0.001     & 0.025     & 0.060  & 0.953            \\
WP        & -0.002    & 0.002     & -0.820 & 0.415            \\
OCH       & 0.003     & 0.005     & 0.520  & 0.600            \\
PWHI      & 0.014     & 0.006     & 2.360  & 0.018            \\
LF        & 0.039     & 0.015     & 2.700  & 0.007            \\
L\_POPDEN & 0.030     & 0.024     & 1.240  & 0.213            \\
CASES     & -0.000001 & 0.000     & -0.180 & 0.860            \\
PR        & 0.051     & 0.020     & 2.530  & 0.011            \\
MHHI      & -0.00001  & 0.000     & -2.130 & 0.033            \\
GR        & 0.001     & 0.0003    & 2.230  & 0.026           \\
\bottomrule
\end{tabular}
\end{table}

\section{Discussions and Conclusions}\label{Sect:Discussions and Conclusions}
Results from the binary logit model explain that tweet width, people living in the western regions of the US, working-class people, gross household rent, and low-income people are positively associated with reopening the economy. On the other hand, average family size and household income are negatively associated with reopening sentiment. However, they have a limited impact. Moreover, the number of Coronavirus cases in a region does not appear to have a significant sentiment association with reopening the economy.

\subsection{Implications}
Several policy recommendations can be drawn from the analysis.
By investigating the socioeconomic factor associations for reopening sentiments, this study implicitly pointed out which strata of the people, and which regions bear positive and negative sentiments for reopening the economy. Thus, the federal and state governments and agencies can take necessary decisions based on the study findings where to intervene to secure people and economy of the country. The findings can also be used to improve information dissemination, as information categories can impact performance, and increase effectiveness of other preventive measures such as antiseptic and disinfectant protocol (e.g., hand washing, body and nasal spray) are testified to reduce infection of the people \cite {samuel2017effects, miller2020blueprint}. Real-time COVID-19 incidence and socioeconomic characteristics of the people provide essential directions to the policymakers and health professionals to allocate resources for developing vaccines and therapeutics to protect people \cite {angulo2020reopening}. Thus, adequate protective actions need to be undertaken for anticipated risks and pave the way to a new normal reopening.

\subsection{Limitations}
Despite timely contribution to the literature, this study has some limitations. First, using tweeter data does not represent the complete section of the population. Very few people use Tweeter \cite {nagar2014case, wang2016spatial, widener2014using}. Thus, the study using Twitter data unable to provide a general idea about the subject matter. Second, sentiment analysis is unable to pick up nuanced or ambiguous meanings (e.g., slang, misspellings, nuanced or ambiguous meanings, Twitter lexicon, inside references, current events, intention, mood) of the tweets which give misleading information \cite {nagar2014case, widener2014using}. Third, it is very difficult to analyze and identify valuable information from a very large quantity of unstructured and heterogeneous data from social media to acquire useful information for decision making \cite {bello2016social}. Fourth, socioeconomic and household information is averaged at the state level which provides a little variation. Thus, a study with fine geographic resolutions (e.g., county, zip code etc.) might provide more insights. 

\subsection{Future research and conclusion}
This research uses a novel methodological variation by combining sentiment analytics from Twitter data, with a custom selection of socioeconomic variables from Census data, to create insights that can contribute to developing a clearer understanding of the factors driving post-COVID-19 reopening sentiment. Sentiment and human behavior can be affected by a wide range of factors, including the information propagation formats, and future research could therefore include relevant time-matched news articles and responses to the Tweets data for sentiment analysis \cite{saif2016contextual, samuel2017information, liu2012emoticon}. This study opens up a valuable stream of research in identifying factors contributing to post-crisis public sentiment using sentiment analysis and can influence future research in policy formation, public mental health, information systems and applications of sentiment analytics. In summary. this study provides interesting insights to researchers and policymakers, and makes two key contributions:  a) the study makes a novel methodological contribution by combining sentiment analytics using Twitter data with Census data, for socioeconomic analysis which can be used for further research, and b) the study provides post-COVID-19 reopening insights into positive sentiment and negative sentiment population segmentations, which can be useful for focused and effective communication and policy management.







\bibliographystyle{elsarticle-num}
\bibliography{Bibliography.bib}

\end{document}